# Unified Growth Theory Contradicted by the Absence of Takeoffs in the Gross Domestic Product

Ron W Nielsen[1]

Environmental Futures Research Institute, Gold Coast Campus, Griffith University, Qld, 4222, Australia

January, 2016

Data describing historical economic growth are analysed. They demonstrate convincingly that the takeoffs from stagnation to growth, claimed in the Unified Growth Theory, never happened. This theory is again contradicted by data, which were used, but never properly analysed, during its formulation. The absence of the claimed takeoffs demonstrates also that the postulate of the differential takeoffs is contradicted by data.

**Introduction**

In our earlier study (Nielsen, 2015a) we have presented mathematical analysis of the historical Gross Domestic Product (GDP). Now, we shall focus our attention on the alleged takeoffs from stagnation to growth, postulated in the Unified Growth Theory (Galor, 2005a, 2008a, 2011, 2012a) and on the associated claim of the differential takeoffs, i.e. on the claim that takeoffs happened at distinctly different time for developed and less-developed regions. "The take-off of developed regions from the Malthusian Regime was associated with the Industrial Revolution and occurred at the beginning of the 19th century, whereas the take-off of less developed regions occurred towards the beginning of the 20th century and was delayed in some countries well into the 20th century" (Galor, 2005a, p. 185). We shall demonstrate that the postulated takeoffs never happened and consequently that the concept of the differential takeoffs is contradicted by data.

The data we are using (Maddison, 2010) are virtually the same (Maddison, 2001) as used by Galor during the formulation of his Unified Growth Theory. The difference between the two compilations is that the new set of data was extended to the 21st century. The extended data are not essential for testing the Unified Growth Theory but they indicate more clearly the continuing transitions to slower trajectories, the process which commenced towards the end of the 20th century.

Galor had access to Maddison's data but he has never analysed them. His interpretations of the mechanism of economic growth are based on strongly questionable quotations of isolated numbers, on the unfortunate simplistic and self-misleading examination of data and on the habitual use of grossly distorted diagrams (Ashraf, 2009; Galor, 2005a, 2005b, 2007, 2008a, 2008b, 2008c, 2010, 2011, 2012a, 2012b, 2012c; Galor and Moav, 2002; Snowdon & Galor, 2008).

---

[1] AKA Jan Nurzynski, r.nielsen@griffith.edu.au; ronwnielsen@gmail.com; http://home.iprimus.com.au/nielsens/ronnielsen.html

Earlier study (Nielsen, 2014) indicated that historical economic growth can be described using hyperbolic distributions in much the same way as the growth of human population (von Foerster, Mora & Amiot, 1960). More recently (Nielsen, 2015a), we have demonstrated that the same description is applicable also to the regional economies.

Unlike the better-known exponential growth, which is easier to understand, hyperbolic distributions are strongly deceptive because they appear to be made of two distinctly different components, slow and fast, joined perhaps by a certain transition component. This illusion is so strong that even the most experienced researchers can be easily deceived particularly if their research is based on a limited body of data, as it was in the past. Fortunately, Maddison's data solve this problem, and fortunately also their analysis is trivially simple because, as pointed out earlier (Nielsen, 2014), hyperbolic distributions can be easily identified and analysed using the reciprocal values of data. Consequently, if in the past, researchers were basing their conclusions on the strongly-limited sets of data and imagined that there was a prolonged epoch of stagnation followed by a sudden takeoff, now there is no excuse to continue with such interpretation of the historical economic growth because we have excellent sets of data, which lead to entirely different interpretations. It is, therefore surprising, if not disappointing, that Galor, who had access to these excellent data and even used them during the formulation of his theory, did not analyse them properly but followed the traditional and incorrect interpretations of the historical economic growth.

Theories play an important role in scientific research because they crystallise interpretations of studied phenomena. However, theories have to be always tested by data. In science it is important to look for data confirming theoretical explanations but it is even more important to discover contradicting evidence, because data confirming a theory confirm only what we already know but contradicting evidence may lead to new discoveries.

Currently, the most complete theory describing the mechanism of the historical economic growth is the Unified Growth Theory (Galor, 2005a, 2008a, 2011, 2012a). One of the fundamental postulates of this theory is the postulate of the existence of three regimes of growth governed by three distinctly different mechanisms: (1) the Malthusian regime of stagnation, (2) the post-Malthusian regime, and (3) the sustained-growth regime. This fundamental postulate is used repeatedly throughout the narrative of the Unified Growth Theory and serves as the essential support for the discussed interpretations and explanations. If this corner stone is removed, the whole structure becomes unsupported.

According to Galor (Galor, 2008a, 2012a), Malthusian regime of stagnation was between 100,000 BC and AD 1750 for developed regions and between 100,000 BC and AD 1900 for less-developed regions. The claimed starting time of this regime appears to be based entirely on conjecture because Maddison's data are terminated at AD 1 and even they contain significant gaps below AD 1500. The post-Malthusian regime was allegedly between AD 1750 and 1850 for developed regions and from 1900 for less-developed regions. The sustained-growth regime was supposed to have commenced around 1850 for developed regions.

The alleged transition at the end of the postulated regime of Malthusian stagnation for various regions and countries is described by Galor as "the sudden take-off from stagnation to growth" (Galor, 2005a, pp. 177, 220, 277), as a "sudden spurt" (Galor, 2005a, 177, 220) or as "remarkable" or "stunning" escape from the Malthusian trap (Galor, 2005a, pp. 177, 220). It is a signature, which cannot be missed.

For developed regions, this signature is supposed to have coincided with the onset of the Industrial Revolution, 1760-1840 (Floud & McCloskey, 1994). "The take-off of developed regions from the Malthusian Regime was associated with the Industrial Revolution" (Galor, 2005a, p. 185). Indeed, the Industrial Revolution is considered to have been "the prime engine of economic growth" (Galor, 2005a, p. 212).

The signature of the takeoffs is characterised by three features: (1) it should be a prominent change in the pattern of growth, (2) it should be a transition from stagnation to growth and (3) it should occur at the time claimed by the theory. For developed regions, the postulated takeoffs should occur around AD 1750. For less-developed regions, they should occur around 1900. Takeoffs serve as a convenient test of the Unified Growth theory not only because they should be prominent but also because there are no significant gaps in the data around the time of their postulated presence and consequently the transition from stagnation to growth should be easily identifiable.

A transition from growth to growth is *not* a signature of the postulated takeoff from stagnation to growth. Thus, for instance, a transition from hyperbolic growth to another hyperbolic growth is not a signature of the sudden takeoff from stagnation to growth. Likewise, a transition at a distinctly different time is not a confirmation of the theoretical expectations.

In a series of earlier publications (Nielsen, 2015c, 2015d, 2015e, 2015f, 2015g, 2016) we have already demonstrated that the concept of the three regimes of growth is contradicted repeatedly by data. In particular, in one of the earlier publications (Nielsen, 2015c), we have demonstrated that this concept is contradicted not only by the GDP data but also by the data describing income per capita (GDP/cap). Consequently, this fundamental corner stone of the Unified Growth Theory has been already undermined.

In the present discussion, we shall show that the existence of the postulated takeoffs from stagnation to growth, which also form an important structural support for this theory, is contradicted by data. The absence of the takeoffs eliminates also another doctrine used in the Demographic Transition Theory, the doctrine of differential takeoffs.

In the future we shall demonstrate that "The mind-boggling phenomenon of the Great Divergence" (Galor, 2005a, p. 220) is mind-boggling only because it is hard to understand how anyone familiar with mathematics could be puzzled by such an artificially-created structure. If hyperbolic distributions are not properly analysed they can be used to generate such phantom and totally meaningless features. It can be also shown that Galor's understanding of the growth rate of income per capita is incorrect.

With so many incorrect concepts and with even a greater number of interpretations based on so many incorrect concepts (one incorrect interpretation leading to another) it is hard to see how much can be redeemed from this theory, which appears to be fundamentally wrong. The best solution would be to replace it by a theory based on the scientific analysis of data.

Throughout the analysis presented here, the values of the Gross Domestic Product (GDP) will be expressed in billions of the 1990 International Geary-Khamis dollars.

**World economic growth**

Results of mathematical analysis of the world economic growth are presented in Figure 1. If the Unified Growth Theory (Galor, 2005a, 2008a, 2011, 2012a) is correct, we should expect to see clear signs of *two* takeoffs: around 1750 for developed regions and around 1900 for less-developed regions. We see none of them.

The data and their analysis are in direct contradiction of this theory. They show that the economic growth was remarkably stable and that the claimed or wished-for takeoffs never happened. The absence of the two claimed takeoffs is strikingly conspicuous. Galor's claim of the "spectacular" or "stunning" escapes from Malthusian trap (Galor, 2005a, pp. 177, 220) is spectacularly and stunningly contradicted by the analysis of the economic-growth data, the same data, which he used, but never properly analysed, during the formulation of his theory.

The absence of the takeoffs has been also demonstrated for the income per capita data (GDP/cap) for the world economic growth (Nielsen, 2015c). In science, such results would have been sufficient to show that the Unified Growth Theory needs to be revised to bring it in agreement with data, however, when closely analysed this theory is found to be repeatedly contradicted by data (Nielsen, 2015a, 2015b, 2015c, 2015d, 2015e, 2015f, 2015g, 2016). The already published discussions represent only a small part of contradicting evidence.

Hyperbolic growth of the world economy is in harmony with the hyperbolic growth of the world population (Nielsen, 2015b; von Foerster, Mora & Amiot, 1960). In both cases, the growth was indeed slow over a long time and fast over a short time. In both cases the growth creates an illusion of stagnation followed by a sudden takeoff. However, in both cases the growth was hyperbolic. There was no stagnation and no sudden takeoff. Furthermore, in both cases the growth started to be diverted, relatively recently, to slower trajectories.

**Western Europe**

The growth of the GDP in Western Europe is shown in Figure 2. Results of analysis show that there was no takeoff from stagnation to growth because (1) there was no stagnation and (2) because the economic growth, which is described well by the hyperbolic trajectory, was stable during the time of the alleged takeoff. The takeoff simply did not happen.

The claim of the stunning or remarkable takeoff is contradicted by data. There was no takeoff of any kind, stunning or less stunning, remarkable or less remarkable, sudden or gradual; none at all. The Industrial Revolution, the alleged "prime engine of economic growth" (Galor, 2005a, p. 212), made no impression on changing the economic growth trajectory in the region where this engine should have been working most efficiently. Industrial Revolution brought many other important changes but, surprisingly perhaps, did not change the economic growth trajectory in the countries closest to this monumental development.

**Eastern Europe**

The analysis of the historical data for Eastern Europe is summarised in Figure 3. There was no stagnation and no takeoff at any time. Industrial Revolution had no impact on changing the economic growth trajectory in the countries of Eastern Europe.

**Former USSR**

The analysis of the data for the countries of the former USSR is presented in Figure 4. There was no stagnation and no takeoff at any time. Industrial Revolution had no impact on changing the economic growth trajectory in the countries of former USSR.

**Asia**

Analysis of the historical economic growth in Asia is summarised in Figure 5. Asia is made primarily of less-developed countries (BBC, 2014; Pereira, 2011) and consequently, according to the Unified Growth Theory (Galor, 2005a, 2008a, 2011, 2012a), economic growth in this region should have been stagnant until around 1900, the year marking the alleged stunning escape from Malthusian trap, the escape manifested by the postulated dramatic takeoff.

The data and their analysis show that there was no stagnation and no claimed takeoff from stagnation to growth. The data reveal a steadily increasing and stable hyperbolic growth until around 1950. From around that year, economic growth *was* diverted to a faster trajectory. This boosting occurred close to the time of the postulated takeoff from stagnation to growth. However, it was *not* a transition from stagnation to growth but from growth to growth.

This change in the growth trajectory was the commonly-observed transition to a slower trajectory but in this case it was preceded by a minor and temporary boosting. It would be interesting to explore and explain this minor boosting but we shall not find its explanation in the Unified Growth Theory. This theory does not even notice this feature.

**Africa**

Results of analysis are presented in Figure 6. Africa is also made of less-developed countries (BBC, 2014; Pereira, 2011) so according to the Unified Growth Theory (Galor, 2005a, 2008a, 2011, 2012a) it should have experienced stagnation in the economic growth until around 1900 followed by a clear takeoff from stagnation to growth around that year. These expectations are contradicted by the economic growth data because (1) economic growth was not stagnant but hyperbolic (Nielsen, 2015d), (2) there was no takeoff from stagnation to growth around 1900 or around any other time close to that year and (3) shortly after the expected time of the takeoff, economic growth in Africa started to be diverted to a slower trajectory.

As discusses elsewhere (Nielsen, 2015d), there was an acceleration in the economic growth in Africa around 1820. However, this acceleration occurred significantly earlier and it was not a transition from stagnation to growth but from growth to growth. Even more specifically, it was a transition from the hyperbolic growth to another hyperbolic growth. It was also acceleration at a wrong time, not around 1900 but around the time of the Industrial Revolution. This acceleration can be explained by noticing that it appears to coincide with the intensified colonisation of Africa (Duignan & Gunn, 1973; McKay, Hill, Buckler, Ebrey, Beck, Crowston, & Wiesner-Hanks, 2012; Pakenham, 1992). The fast increasing GDP after 1820 was not reflecting the rapidly improving living conditions of African population brought about by the beneficial changes caused by the Industrial Revolution but the rapidly increasing wealth of new settlers and their countries of origin at the expense of the deploring living conditions of the native populations.

The takeoff from stagnation to growth, claimed by the Unified Growth Theory (Galor, 2005a, 2008a, 2011, 2012a), did not happen in the region where it should have been prominently present. Economic growth was always stable in Africa (Nielsen, 2015d) and now it is being diverted to a slower trajectory. Escape from the Malthusian trap never happened because there was no trap.

**Latin America**

Results of the analysis of the economic growth in Latin America are presented in Figure 7. Latin America is also made of less-developed countries (BBC, 2014; Pereira, 2011) so again, according to the Unified Growth Theory (Galor, 2005a, 2008a, 2011, 2012a), economic growth in this regions should have been stagnant until around 1900 and fast-increasing from around that year. This pattern of growth is stunningly contradicted by data, the same data, which were used, but never properly analysed, during the formulation of this theory. At the time of the claimed "stunning" and "remarkable" escape from Malthusian trap (Galor, 2005a, pp. 177, 220) economic growth in Latin America was already diverted to a slower trajectory.

**Summary and conclusions**

Results of mathematical analysis of Maddison's data (Maddison, 2010) show convincingly that takeoffs from stagnation to growth, claimed repeatedly in the Unified Growth Theory (Galor, 2005a, 2008a, 2011, 2012a) never happened. The growth of the GDP was not stagnant but hyperbolic and, in general, remarkably stable.

Galor used the earlier compilation of Maddison's data (Maddison, 2001) but the new compilation contains the same data until the end of the 20th century. The claimed takeoffs were supposed to have occurred during the 20th century so any of these compilations can be used to check the Unified Growth Theory. This theory is again contradicted by the same data, which were used, but never properly analysed, during its formulation.

We have demonstrated that there were no takeoffs for the world economic growth and no takeoffs in Western Europe, Eastern Europe, Asia, in countries of the former USSR, in Africa and in Latin America. The analysis presented here and the earlier studies (Nielsen, 2015a, 2015b, 2015c, 2015d, 2015e, 2015f, 2015g, 2016) show that concepts of prolonged stagnation followed by a "remarkable" or "stunning" escape from Malthusian trap (Galor, 2005a, pp. 177, 220) are contradicted by data.

In science, such overwhelming evidence would have been more than sufficient to show that the theory is unacceptable and that it should be either thoroughly revised or rejected and replaced by a more suitable theory, a theory based on a scientific analysis of data, a reliable theory, which could be used in the economic growth research. In its present form, Unified Growth Theory is neither reliable nor useful. In fact it is strongly misleading.

Our analysis of Maddison's data (Maddison, 2010) shows not only that the concept of Malthusian regime of stagnation followed by dramatic escapes from Malthusian trap is incorrect but also that the concept of the differential takeoffs is incorrect because we cannot have differential takeoffs without takeoffs.

Unified Growth Theory is riddled with questionable claims and interpretations. In due time, we shall demonstrate that this theory is contradicted by regional GDP/cap data in much the same way as it is contradicted by the global data (Nielsen, 2015c). We shall show that this theory is contradicted by the economic growth in the UK, the centre of the Industrial Revolution where the Unified Growth Theory should have the strongest support. It can be also shown that this theory is contradicted by the economic growth in other individual countries.

We shall demonstrate that the postulate of the great divergence is also based on the incorrect interpretation of the mathematical properties of hyperbolic distributions. Furthermore, we shall demonstrate that Galor's repeated interpretation of growth rates of income per capita is incorrect.

In its present form, Unified Growth Theory is unacceptable. In order to improve it, it would be necessary to examine it closely to determine not only how much of it is based on the incorrect interpretation of data but also how much is just a pure fantasy. However, the best solution would probably be to replace it by a new theory.

**World Economic Growth**

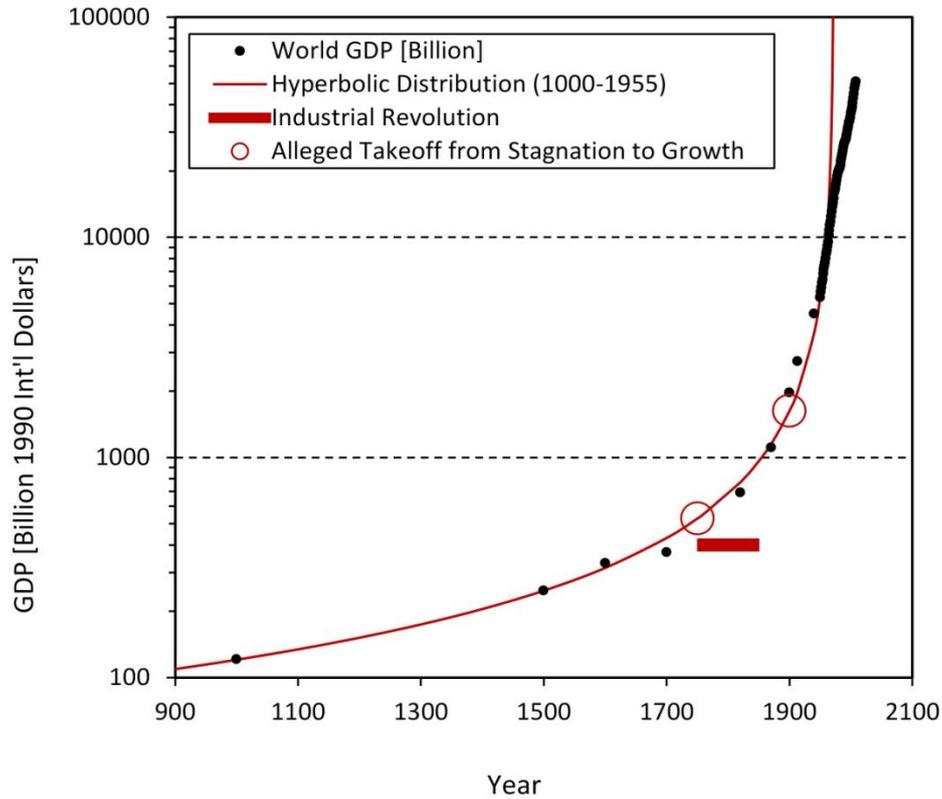

**Figure 1.** No takeoffs from stagnation to growth. Two postulated takeoffs are indicated (Galor, 2005a, 2008a, 2011, 2012a): for developed regions around 1750 and for less-developed regions around 1900. The world economic growth was not stagnant but hyperbolic and it was remarkably stable. Industrial Revolution, "the prime engine of economic growth" (Galor, 2005a, p. 212), had no impact on changing the economic growth trajectory. Unified Growth Theory (Galor, 2005a, 2008a, 2011, 2012a) is contradicted by data. For further discussion of the world economic growth see Nielsen (2015a, 2015c).

## Western Europe

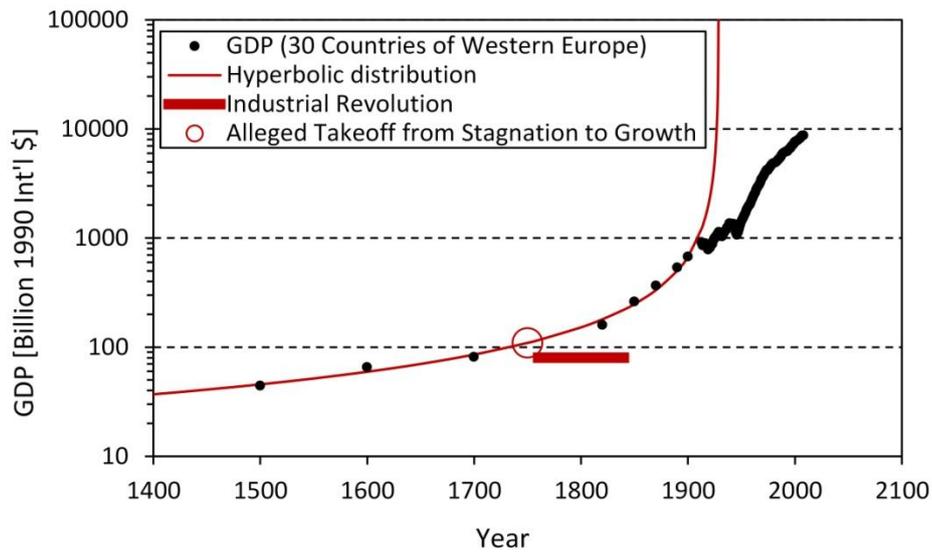

**Figure 2.** No takeoff from stagnation to growth. Economic growth in Western Europe was not stagnant but hyperbolic and it was remarkably stable. Industrial Revolution, "the prime engine of economic growth" (Galor, 2005a, p. 212), had no impact on changing the economic growth trajectory where this "engine" should have worked most efficiently. Unified Growth Theory (Galor, 2005a, 2008a, 2011, 2012a) is contradicted by data. For further discussion of economic growth in Western Europe see Nielsen (2015g)

**Eastern Europe**

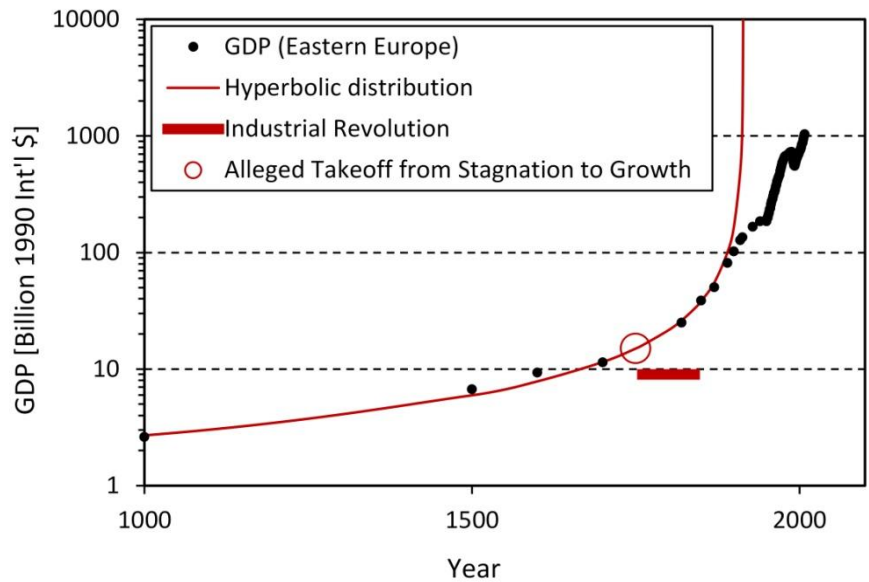

**Figure 3**. No takeoff from stagnation to growth. Economic growth in Eastern Europe was not stagnant but hyperbolic and it was remarkably stable. Industrial Revolution, "the prime engine of economic growth" (Galor, 2005a, p. 212), had no impact on changing the economic growth trajectory. Unified Growth Theory (Galor, 2005a, 2008a, 2011, 2012a) is contradicted by data. For further discussion of economic growth in Eastern Europe see Nielsen (2015g)

**Former USSR**

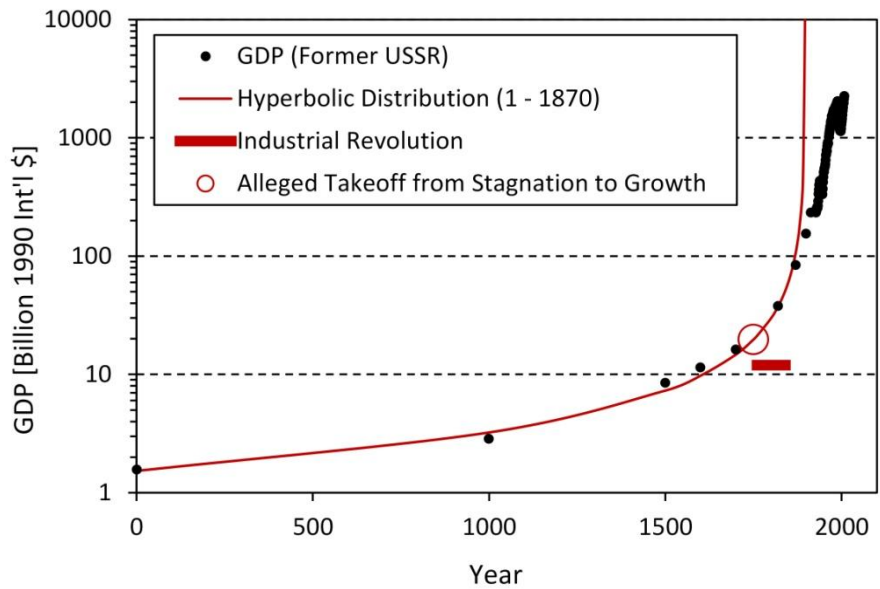

**Figure 4.** No takeoff from stagnation to growth. Economic growth in the former USSR was not stagnant but hyperbolic and it was remarkably stable. Industrial Revolution, "the prime engine of economic growth" (Galor, 2005a, p. 212), had no impact on changing the economic growth trajectory. Unified Growth Theory (Galor, 2005a, 2008a, 2011, 2012a) is contradicted by data. For further discussion of economic growth in the former USSR see Nielsen (2015f)

**Asia**

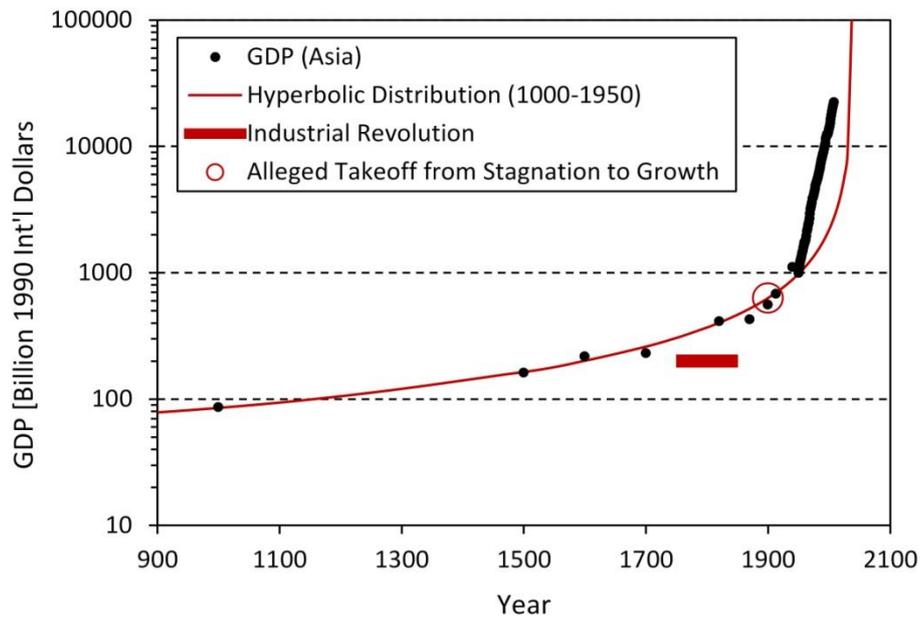

**Figure 5.** No takeoff from stagnation to growth. Economic growth in Asia was not stagnant but hyperbolic before the alleged takeoff and it was remarkably stable. The minor boosting after the alleged takeoff was not a transition from stagnation to growth but a transition from growth to growth. It was similar to the commonly-observed transitions to slower trajectories but in this case it was preceded by a minor and temporary boosting. Unified Growth Theory (Galor, 2005a, 2008a, 2011, 2012a) is contradicted by data. For further discussion of economic growth in Asia see Nielsen (2015e).

**Africa**

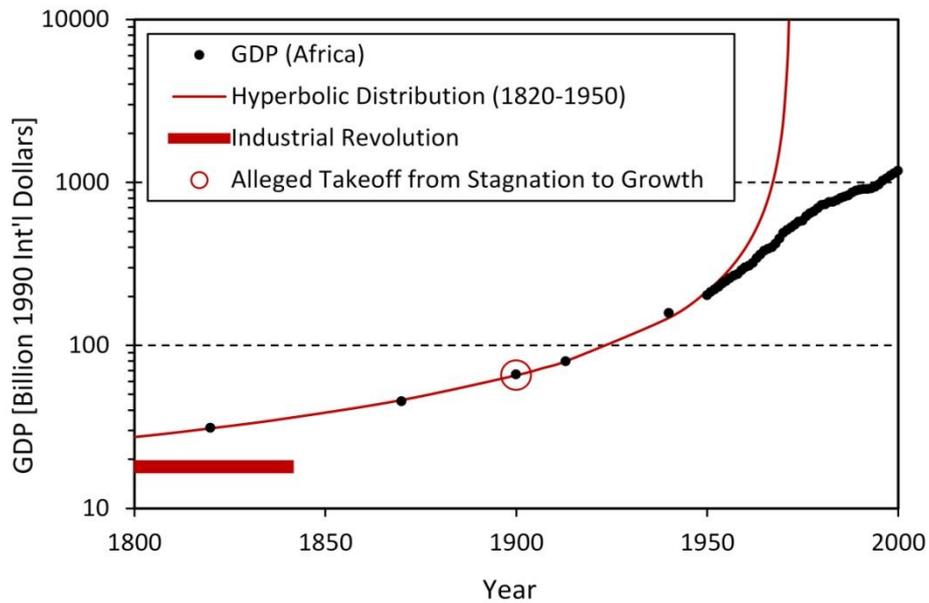

**Figure 6.** No takeoff from stagnation to growth. Economic growth in Africa was not stagnant but hyperbolic. Unified Growth Theory (Galor, 2005a, 2008a, 2011, 2012a) is contradicted by data. Shortly after the alleged dramatic but non-existent escape from the postulated Malthusian trap, economic growth in Africa started to be diverted to a slower trajectory. For further discussion of economic growth in Africa see Nielsen (2015d).

**Latin America**

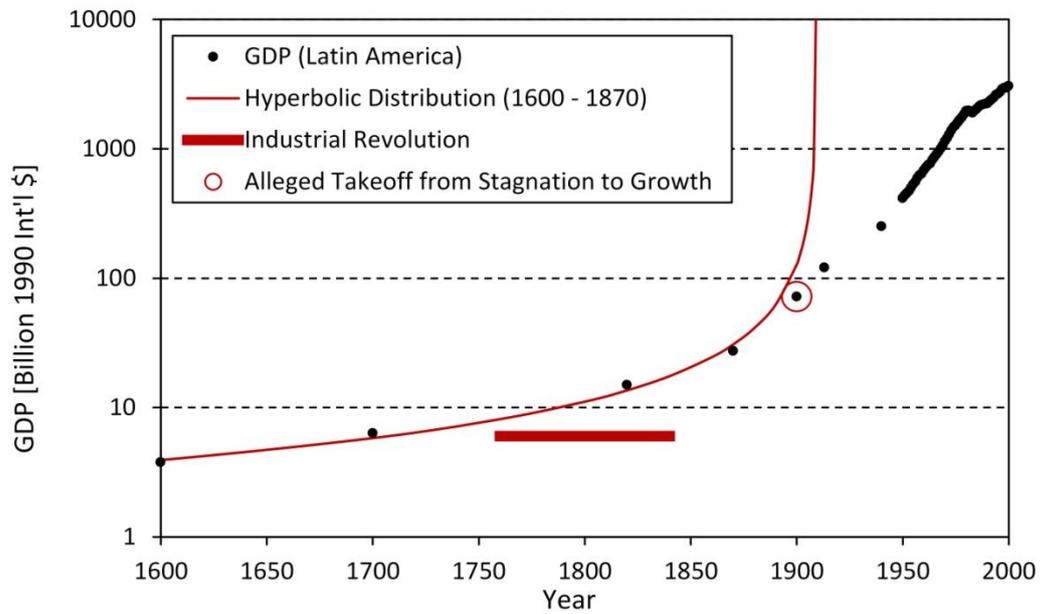

**Figure 7.** No takeoff from stagnation to growth. Economic growth in Latin America was not stagnant but hyperbolic. At the time of the alleged takeoff, economic growth in Latin America was already following a slower trajectory. The alleged takeoff is replaced by a slower growth. The "spectacular" or "stunning" escapes from Malthusian trap (Galor, 2005a, pp. 177, 220) never happened. Unified Growth Theory (Galor, 2005a, 2008a, 2011, 2012a) is contradicted by data. For further discussion of economic growth in Latin America see Nielsen (2016).